\documentclass[]{spie}
\pdfoutput=1
\usepackage{graphicx, color}
\usepackage[colorlinks=true]{hyperref}

\usepackage{subcaption}
\usepackage{booktabs, tabularx}
\usepackage{amsmath,amsfonts,amssymb}
\usepackage[table]{xcolor}

\definecolor{brown}{rgb}{0.66,0.33,0 }

\def\papertitle{Weak Supervision in Convolutional Neural Network for Semantic Segmentation of Diffuse Lung Diseases Using Partially Annotated Dataset}
\def\paperkeywords{deep learning, weak supervision, convolutional neural network, diffuse lung diseases, semantic segmentation}
\def\github{https://github.com/yk-szk/SPIE2020}

\title{\papertitle}

\author[a]{Yuki Suzuki}
\author[a]{Kazuki Yamagata}
\author[a]{Yanagawa Masahiro}
\author[a]{Shoji Kido}
\author[a]{Noriyuki Tomiyama}
\affil[a]{Graduate School of Medicine, Osaka University, 2-2 Yamadaoka, Suita, Osaka, 565-0871, Japan}
\authorinfo{Further author information: (Send correspondence to Yuki Suzuki)\\
  Yuki Suzuki: E-mail: y-suzuki@radiol.med.osaka-u.ac.jp, Telephone: \texttt{+}81 (0)6 6879 3432\\
  Shoji Kido: E-mail: kido@radiol.med.osaka-u.ac.jp, Telephone: \texttt{+}81 (0)6 6879 3432}

\hypersetup{
  pdftitle={\papertitle},
  pdfauthor={Yuki Suzuki},
  pdfkeywords={\paperkeywords}
}

\begin{document}
\maketitle

\begin{abstract}
  Computer-aided diagnosis system for diffuse lung diseases (DLDs) is necessary for the objective assessment of the lung diseases.
  In this paper, we develop semantic segmentation model for 5 kinds of DLDs.
  DLDs considered in this work are consolidation, ground glass opacity, honeycombing, emphysema, and normal.
  Convolutional neural network (CNN) is one of the most promising technique for semantic segmentation among machine learning algorithms.
  While creating annotated dataset for semantic segmentation is laborious and time consuming, creating partially annotated dataset, in which only one chosen class is annotated for each image, is easier since annotators only need to focus on one class at a time during the annotation task.
  In this paper, we propose a new weak supervision technique that effectively utilizes partially annotated dataset.
  The experiments using partially annotated dataset composed 372 CT images demonstrated that our proposed technique significantly improved segmentation accuracy.
\end{abstract}


\section{INTRODUCTION}
Diffuse lung diseases (DLDs) are pulmonary abnormalities observable in medical images such as chest computed tomography (CT).
Computer-aided diagnosis system for diffuse lung diseases is necessary to eliminate interobserver variabilities\cite{Watadani2013} and achieve objective assessment of DLDs, that can lead to better diagnosis and patient treatment.
Therefore, our goal was to develop an automated DLD segmentation method for the objective assessment of DLDs.
The DLD patterns considered in this paper are consolidation (CON), ground glass opacity (GGO), honeycombing (HCM), emphysema (EMP), and normal (NOR).

A number of studies regarding automated assessment of DLDs have been conducted in various context including image level classification and semantic segmentation.
There are several kinds of supervised methods for automated assessment of DLDs including fully-supervised\cite{Hashimoto2018,gao2018holistic,Negahdar2019}, semi-supervised\cite{Anthimopoulos2019}, weakly supervised\cite{Wang2019}, and unsupervised\cite{Mabu2017}.
Machine learning techniques are widely used for semantic segmentation since they are capable of learning complicated texture patterns and often outperform hand-crafted algorithms.
Among the machine learning techniques, convolutional neural network (CNN) is one of the most  successful technique in computer vision tasks.
One of the biggest drawbacks of machine learning including CNN is that it requires training dataset, which typically involves costly annotation tasks unless there is publicly available annotated dataset.
Ideal annotation for semantic segmentation is pixel-wise full annotation, in which every pixel in the image is labeled as one of the possible classes.

In this paper, we define partial annotation as an annotation format in which only one class is chosen for the annotation and only pixels belonging to the chosen class are annotated per image.
For example, in Figure \ref{fig:example_con}, although there is ground glass opacity in the image, only consolidation is chosen for annotation and pixels of consolidation are annotated.
Partially annotated dataset is less informative for training, however, it is much easier to create compared to fully annotated dataset since annotators only need to focus on one class at a time during the annotation task.

Partially annotated datasets have been utilized previously \cite{Dmitriev_2019_CVPR,Kong2019}.
In this paper, we propose a new weak supervision technique that fully utilizes partially annotated dataset.
Throughout this paper, each DLD pattern is represented or painted in the following colors (CON:cyan, GGO:yellow, HCM:red, EMP:green, NOR:brown.)

\section{MATERIAL AND METHOD}
\subsection{Dataset}
The dataset used in this study consists of 372 high-resolution computed tomography (HRCT) taken in Yamaguchi University Hospital, Japan.
The mean and the standard deviation of the pixel size are $0.684 mm$ and $0.0517 mm$ respectively and slice thickness is $1 mm$.
In this work, no pixel size equalization was performed since the deviation in the pixel sizes was negligibly small.

\begin{figure}[htbp]
  \centering
  \begin{subfigure}{.16\textwidth}
    \includegraphics[width=1\linewidth]{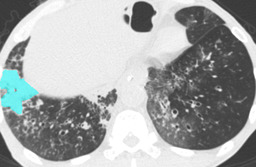}
    \caption{CON \textcolor{cyan}{$\blacksquare$}} \label{fig:example_con}
  \end{subfigure}
  \qquad
  \begin{subfigure}{.16\textwidth}
    \includegraphics[width=1\linewidth]{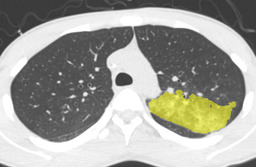}
    \caption{GGO \textcolor{yellow}{$\blacksquare$}}
  \end{subfigure}
  \qquad
  \begin{subfigure}{.16\textwidth}
    \includegraphics[width=1\linewidth]{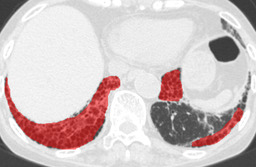}
    \caption{HCM \textcolor{red}{$\blacksquare$}}
  \end{subfigure}
  \qquad
  \begin{subfigure}{.16\textwidth}
    \includegraphics[width=1\linewidth]{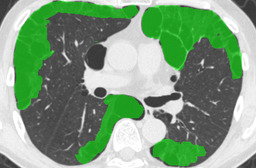}
    \caption{EMP \textcolor{green}{$\blacksquare$}}
  \end{subfigure}
  \qquad
  \begin{subfigure}{.16\textwidth}
    \includegraphics[width=1\linewidth]{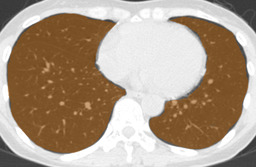}
    \caption{NOR \textcolor{brown}{$\blacksquare$}}
  \end{subfigure}
  \caption{Typical slices for each DLD classes.
    Slices of HRCT are shown in  lung window setting (window-center=-600, window-width=1500) with annotated labels superimposed in transparent colors.
    Note that even if more than one DLD patterns existed, only one DLD pattern was chosen and annotated for a slice to facilitate the annotation process.}
  \label{image_examples}
\end{figure}

Statistics of our dataset are shown in Table \ref{dataset_stats} and typical images and their annotations for each DLD pattern are shown in Figure \ref{image_examples}.
In our partially annotated dataset, all the pixels in a slice were manually classified into two classes: dominating DLD pattern and other tissues.
In other words, all the pixels in our dataset were assigned one of the labels from either of the two label sets, $L_{strong} = \{l_{CON}, l_{GGO}, l_{HCM}, l_{EMP}, l_{NOR}\}$ or $L_{weak} = \{l_{\overline{CON}}, l_{\overline{GGO}}, l_{\overline{HCM}}, l_{\overline{EMP}}, l_{\overline{NOR}}\}$.
For example, in Figure \ref{fig:example_con}, colored pixels were labeled as $l_{CON}$ and all the other pixels were labeled as $l_{\overline{CON}}$.
In this paper, we call pixels of label $l \in L_{weak}$ and $l \in L_{strong}$ as weakly annotated pixels and strongly annotated pixels respectively.
Our pixel-wise annotations were created in the following steps.
First, up to 3 slices were chosen for the annotation for each HRCT scan and for each slice, one representing DLD pattern was chosen by a radiologist.
Second, three radiologists performed pixel-wise binary annotation (e.g. binary annotation between $l_{CON}$ or $l_{\overline{CON}}$) for each slice.
Finally, the radiologists' annotations were merged by taking majority classes for each pixel (i.e. pixels labeled as a DLD pattern by more than 2 radiologists became the corresponding DLD pixel).
In addition to the DLDs annotation, lung fields were manually segmented under the supervision of radiologists and training and testing were conducted only within the lung fields.

\begin{table}[htbp]
  \begin{center}
    \caption{Statistics of the dataset}
    \begin{tabular}{lcccccc}
      \toprule
      \multicolumn{1}{c}{} & CON & GGO & HCM & EMP & NOR & total \\
      \midrule
      \# of pixels ($\times10^5$)& 6 & 16 & 13 & 41 & 25 & 103 \\
      \# of slices & 150 & 114 & 129 & 163 & 55 & 611 \\
      \bottomrule
    \end{tabular}
    \label{dataset_stats}
  \end{center}
\end{table}

\subsection{Method}
U-Net\cite{Ronneberger2015} and its variants are widely used in semantic segmentation of medical images because of its simplicity and applicability.
In this study, we modified U-Net to satisfy two demands for our use.
(1) Take 3D input to leverage 3D spatial information of HRCT.
(2) Generate 2D image since annotations are given by slices not by volumes.
Our modified U-Net's input tensor shape is (6, 512, 512, 1) and output tensor shape is (1, 512, 512, 5), where each axis in the tensor represents z, y, x, and channel respectively.
In the network, the size along z axis was reduced from 6 to 1 by adjusting the padding size of the convolutional layers.
The value 6 in the input tensor shape represents the number of slices that our model takes as the input and it was determined empirically.

\def\vyhat{\mathbf{\hat{y}}} 
The proposed loss function used in the training is shown in Eq. (\ref{loss_function}), where $l$, $\vyhat$, and $H(.)$ denote the ground truth label, the softmax output of the CNN, and cross entropy respectively.
$e(l)$ is one-hot encoding function that works in the conventional way for $l\in L_{strong}$ while for $l\in L_{weak}$, it works so that weakly annotated pixels get encoded equally as the corresponding annotated pixels (e.g. $e(l_{CON}) = e(l_{\overline{CON}})$).
$\lambda$ is the weight for the weakly supervised pixels, which adjust the balance between supervised and weakly supervised pixels.
Our loss function is designed to penalize weakly annotated pixels being predicted as corresponding label.

\begin{equation}
  \mathcal{L}(l,\vyhat)=
  \begin{cases}
    H(e(l),\vyhat), & l \in L_{\text{strong}} \\
    -\lambda H(e(l),\vyhat), & l \in L_{\text{weak}}
  \end{cases}
  \label{loss_function}
\end{equation}

\section{RESULTS AND DISCUSSION}
5-fold stratified cross validation was performed for training and testing.
Stratified method was chosen so that each class of the DLD patterns was equally split into the cross-validation subsets.
In addition to the stratification, during the splitting process, case information was taken into account to avoid data leakage.
During the training, 20\% of slices in the training subset were excluded as validation subset and used for early termination of the training.
Adam optimizer with default parameters was used to train the network.
Proposed method is implemented in Python using Keras library and the source code is publicly available at \href{\github}{\github}.
In this experiment, we compared the following 4 training methods
\{``supervised only'' : base line method that only uses strongly annotated pixel (equivalent of the proposed method with $\lambda=0$) ; ``proposed ($\lambda=0.1$)'' : proposed method with $\lambda=0.1$; ``proposed ($\lambda=1$)'' : proposed method with $\lambda=1$; ``semi-supervised'' : semi-supervised method used by Anthimopoulos, M. et.al.\cite{Anthimopoulos2019}, that utilizes weakly annotated pixels for semi-supervision.\}

Recall, precision, and dice coefficient (a.k.a F-measure) were used for the evaluation.
For the sake of the evaluation, continuous softmax outputs were converted into discrete class labels by selecting the classes that gave the maximum probability.
Table \ref{table:result} shows the evaluated metrics for each method.
By paired t-tests, statistically significant differences were confirmed between the proposed method ($\lambda=0.1$) and other methods in dice coefficients.
As shown in Table \ref{table:result}, utilizing weakly annotated pixels increased precision and $\lambda=0.1$ was the optimal value that balances recall and precision in this experiment.
Evaluated dice coefficients for the proposed method ($\lambda=0.1$) are shown in Figure \ref{dice_swarm}.
As shown in Figure \ref{dice_swarm}, even though the proposed method improved the segmentation accuracy, segmentation accuracy varies between slices.
Figure \ref{confusion_matrix} shows the confusion matrix of the pixel-wise classification result.
In Figure \ref{confusion_matrix}, $L_{weak}$ pixels misclassified as corresponding $L_{strong}$ (e.g. pixels of $l_{\overline{CON}}$ classified as $l_{CON}$) are represented as ``Others''.
As shown in Figure \ref{confusion_matrix}, DLD class combinations with similar texture patterns such as HCM and EMP were misclassified into each other.
Figure \ref{average_results} shows the average result for each DLD class and tested method.

\renewcommand\tabularxcolumn[1]{m{#1}} 
\newcolumntype{Y}{>{\centering\arraybackslash}X} 

\begin{figure}[tbp]
  \begin{center}
    \mbox{}
    \hfill
    \begin{minipage}[c]{.55\textwidth}
      \captionof{table}{The mean values of recall, precision, and dice coefficient for each method.}
      \label{table:result}
      \scriptsize
      \rowcolors{0}{gray!10}{}
      \begin{tabularx}{\linewidth}{YYYY}
        \hiderowcolors
        \toprule
        & Recall & Precision & Dice coefficient \\
        \midrule
        \showrowcolors
        supervised only & $.904 \pm .184$ & $.659 \pm .238$ & $.728 \pm .215$ \\
        proposed ($\lambda=.1$) & $.899 \pm .154$ & $.728 \pm .201$ & $\mathbf{.780 \pm .169}$ \\
        proposed ($\lambda=1$) & $.726 \pm .275$ & $\mathbf{.765 \pm .213}$ & $.706 \pm .246$ \\
        semi-supervised\cite{Anthimopoulos2019} & $\mathbf{.931 \pm .165}$ & $.647 \pm .23$ & $.729 \pm .209$ \\
        \bottomrule
      \end{tabularx}

      \includegraphics[width=\linewidth]{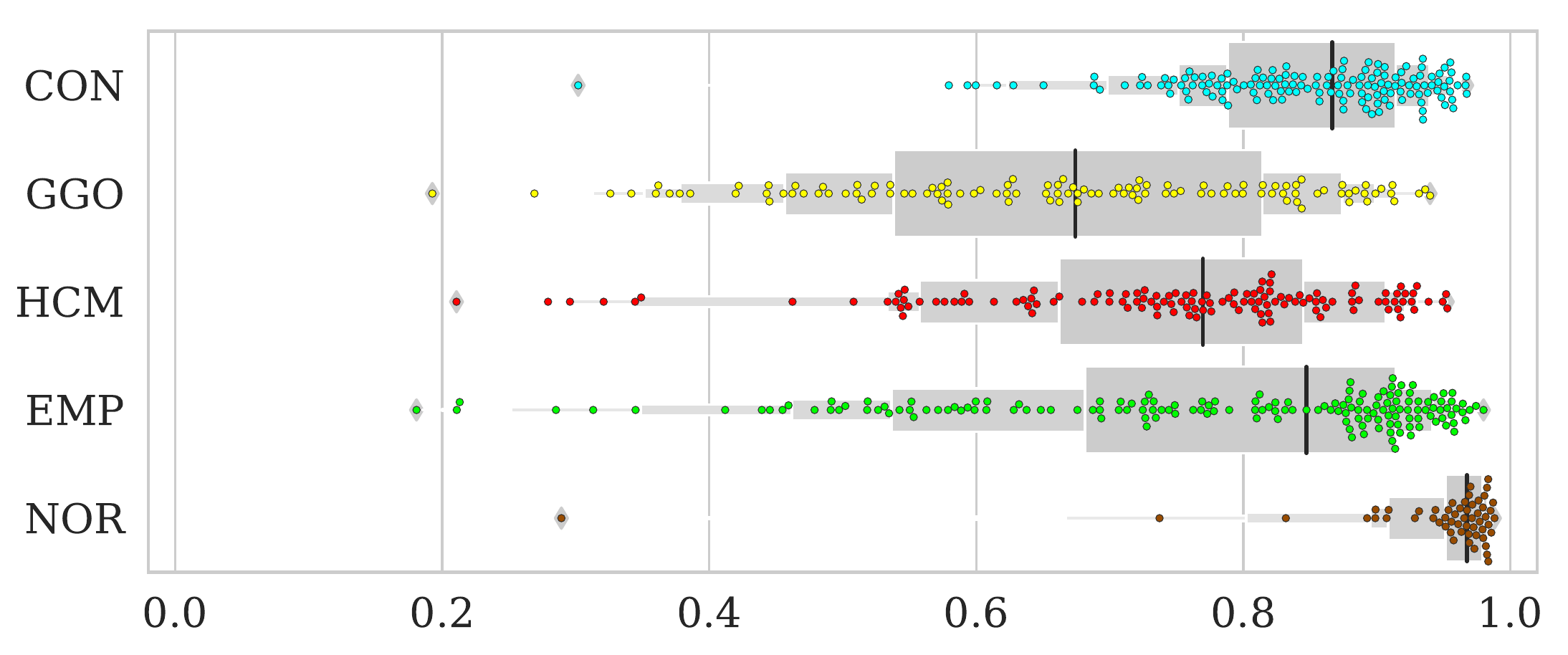}
      \captionof{figure}{Swarm plot on top of a letter value plot of the dice coefficient for the proposed method ($\lambda=0.1$).}
      \label{dice_swarm}
    \end{minipage}
    \hfill
    \begin{minipage}[c]{.4\textwidth}
      \includegraphics[width=\linewidth]{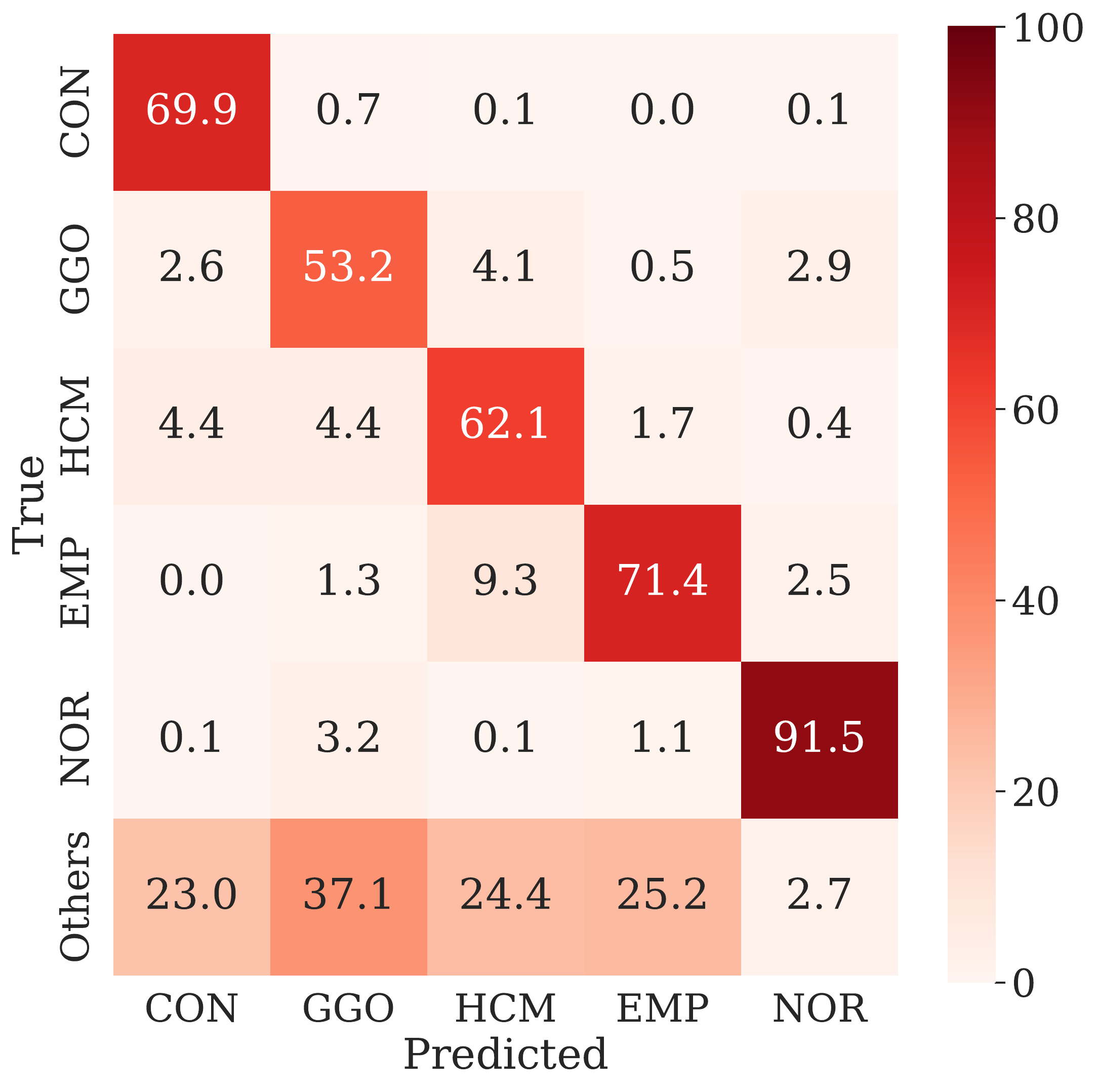}
      \caption{Confusion matrix for the proposed method ($\lambda=0.1$). Values are normalized along Y axis thus diagonal elements indicate the precisions.}
      \label{confusion_matrix}
    \end{minipage}
    \hfill
    \mbox{}
  \end{center}
\end{figure}

\newcommand{\vertical}[1]{\rotatebox[origin=c]{90}{#1}}
\begin{figure}[htbp]
  \begin{tabular}{ccccc}
    & Ground truth & Supervised only &  Proposed ($\lambda=0.1$) & Proposed ($\lambda=1$) \\

    \vertical{CON \textcolor{cyan}{$\blacksquare$}} &
    \begin{minipage}[c]{.21\textwidth}
      \centering
      \includegraphics[width=1\textwidth]{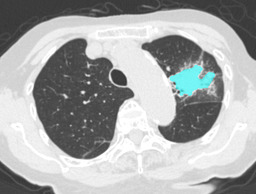}
    \end{minipage} &
    \begin{minipage}[c]{.21\textwidth}
      \centering
      \includegraphics[width=1\textwidth]{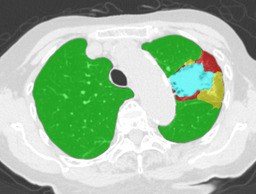}
    \end{minipage} &
    \begin{minipage}[c]{.21\textwidth}
      \centering
      \includegraphics[width=1\textwidth]{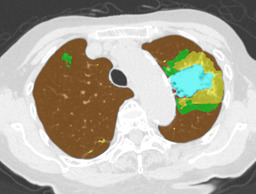}
    \end{minipage} &
    \begin{minipage}[c]{.21\textwidth}
      \centering
      \includegraphics[width=1\textwidth]{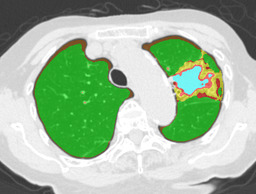}
    \end{minipage}
    \\
    & & 0.839 & 0.868 & 0.824
    \\
    \vertical{GGO \textcolor{yellow}{$\blacksquare$}} &
    \begin{minipage}[c]{.21\textwidth}
      \centering
      \includegraphics[width=1\textwidth]{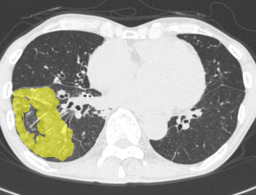}
    \end{minipage} &
    \begin{minipage}[c]{.21\textwidth}
      \centering
      \includegraphics[width=1\textwidth]{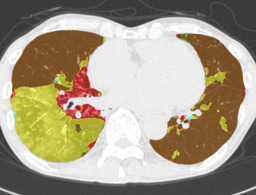}
    \end{minipage} &
    \begin{minipage}[c]{.21\textwidth}
      \centering
      \includegraphics[width=1\textwidth]{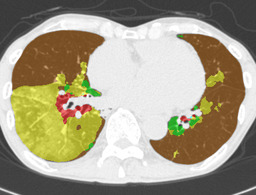}
    \end{minipage} &
    \begin{minipage}[c]{.21\textwidth}
      \centering
      \includegraphics[width=1\textwidth]{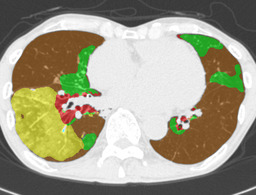}
    \end{minipage}
    \\
    & & 0.693 & 0.676 & 0.876
    \\
    \vertical{HCM \textcolor{red}{$\blacksquare$}} &
    \begin{minipage}[c]{.21\textwidth}
      \centering
      \includegraphics[width=1\textwidth]{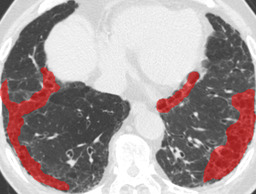}
    \end{minipage} &
    \begin{minipage}[c]{.21\textwidth}
      \centering
      \includegraphics[width=1\textwidth]{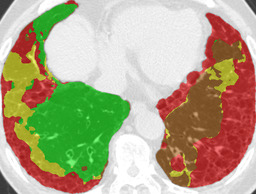}
    \end{minipage} &
    \begin{minipage}[c]{.21\textwidth}
      \centering
      \includegraphics[width=1\textwidth]{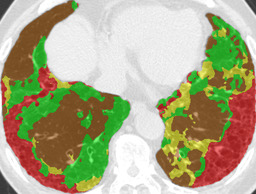}
    \end{minipage} &
    \begin{minipage}[c]{.21\textwidth}
      \centering
      \includegraphics[width=1\textwidth]{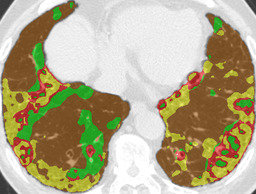}
    \end{minipage}
    \\
    & & 0.581 & 0.770 & 0.435
    \\
    \vertical{EMP \textcolor{green}{$\blacksquare$}} &
    \begin{minipage}[c]{.21\textwidth}
      \centering
      \includegraphics[width=1\textwidth]{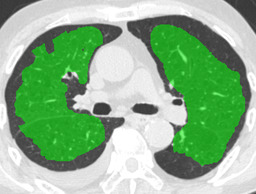}
    \end{minipage} &
    \begin{minipage}[c]{.21\textwidth}
      \centering
      \includegraphics[width=1\textwidth]{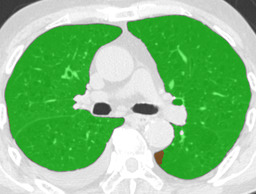}
    \end{minipage} &
    \begin{minipage}[c]{.21\textwidth}
      \centering
      \includegraphics[width=1\textwidth]{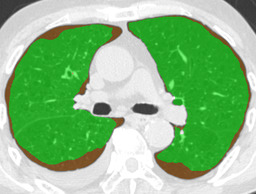}
    \end{minipage} &
    \begin{minipage}[c]{.21\textwidth}
      \centering
      \includegraphics[width=1\textwidth]{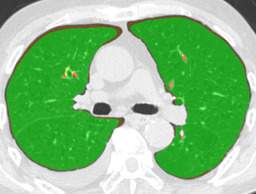}
    \end{minipage}
    \\ & & 0.793 & 0.847 & 0.815
    \\
    \vertical{NOR \textcolor{brown}{$\blacksquare$}} &
    \begin{minipage}[c]{.21\textwidth}
      \centering
      \includegraphics[width=1\textwidth]{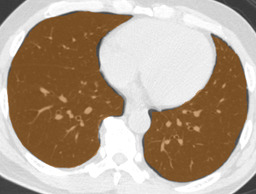}
    \end{minipage} &
    \begin{minipage}[c]{.21\textwidth}
      \centering
      \includegraphics[width=1\textwidth]{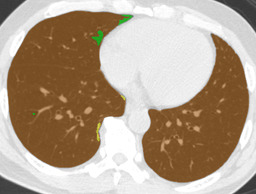}
    \end{minipage} &
    \begin{minipage}[c]{.21\textwidth}
      \centering
      \includegraphics[width=1\textwidth]{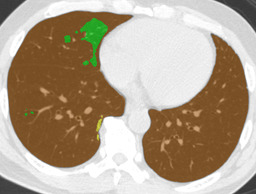}
    \end{minipage} &
    \begin{minipage}[c]{.21\textwidth}
      \centering
      \includegraphics[width=1\textwidth]{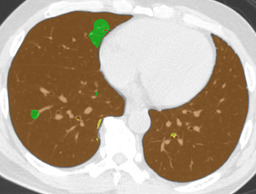}
    \end{minipage}
    \\
    & & 0.978 & 0.968 & 0.974
    \\

  \end{tabular}
  \caption{Average results and dice coefficients for each DLD pattern.
    Automated segmentation results are superimposed with colors.
    For each DLD pattern, the slice that gave the median dice coefficient for the proposed method with $\lambda=0.1$ was chosen to represent the average result.
    Note that although CNN performed multi-class segmentation, only one DLD pattern per slice was taken into account for the evaluation.}
  \label{average_results}
\end{figure}

\section{CONCLUSION}
We proposed a new weakly supervised training  that effectively utilizes weakly annotated pixels of partially annotated dataset.
Experiments demonstrated that our proposed method outperformed conventional methods.
Further work is required to differentiate DLD patterns that have similar texture patterns such as HCM and EMP to improve the segmentation accuracy.

\section*{ACKNOWLEDGEMENTS}
This work was supported by JSPS KAKENHI Grant Number 17H02110.

\bibliographystyle{spiebib}
\bibliography{references}

\begin{thebibliography}{10}

\bibitem{Watadani2013}
Watadani, T., Sakai, F., Johkoh, T., Noma, S., Akira, M., Fujimoto, K.,
  Bankier, A.~A., Lee, K.~S., M{\"{u}}ller, N.~L., Song, J.-W., Park, J.-S.,
  Lynch, D.~A., Hansell, D.~M., Remy-Jardin, M., Franquet, T., and Sugiyama,
  Y., ``{Interobserver Variability in the CT Assessment of Honeycombing in the
  Lungs},'' {\em Radiology}~{\bf 266},  936--944 (mar 2013).

\bibitem{Hashimoto2018}
Hashimoto, N., Suzuki, K., Liu, J., Hirano, Y., MacMahon, H., and Kido, S.,
  ``{Deep neural network convolution (NNC) for three-class classification of
  diffuse lung disease opacities in high-resolution CT (HRCT): Consolidation,
  ground-glass opacity (GGO), and normal opacity},'' in [{\em Medical Imaging
  2018: Computer-Aided Diagnosis}{\nolinebreak\hspace{0.1em}]},  Mori, K. and
  Petrick, N., eds.,  {\bf 10575},  113, SPIE (feb 2018).

\bibitem{gao2018holistic}
Gao, M., Bagci, U., Lu, L., Wu, A., Buty, M., Shin, H.-C., Roth, H., Papadakis,
  G.~Z., Depeursinge, A., Summers, R.~M., and Others, ``{Holistic
  classification of CT attenuation patterns for interstitial lung diseases via
  deep convolutional neural networks},'' {\em Computer Methods in Biomechanics
  and Biomedical Engineering: Imaging {\&} Visualization}~{\bf 6}(1),  1--6
  (2018).

\bibitem{Negahdar2019}
Negahdar, M., Coy, A., and Beymer, D., ``{An End-to-End Deep Learning Pipeline
  for Emphysema Quantification Using Multi-label Learning},'' in [{\em 2019
  41st Annual International Conference of the IEEE Engineering in Medicine and
  Biology Society (EMBC)}{\nolinebreak\hspace{0.1em}]},   929--932, Institute
  of Electrical and Electronics Engineers (IEEE) (oct 2019).

\bibitem{Anthimopoulos2019}
Anthimopoulos, M., Christodoulidis, S., Ebner, L., Geiser, T., Christe, A., and
  Mougiakakou, S., ``{Semantic Segmentation of Pathological Lung Tissue With
  Dilated Fully Convolutional Networks},'' {\em IEEE Journal of Biomedical and
  Health Informatics}~{\bf 23},  714--722 (mar 2019).

\bibitem{Wang2019}
Wang, C., Moriya, T., Hayashi, Y., Roth, H., Lu, L., Oda, M., Ohkubo, H., and
  Mori, K., ``{Weakly-supervised deep learning of interstitial lung disease
  types on CT images},'' in [{\em Medical Imaging 2019: Computer-Aided
  Diagnosis}{\nolinebreak\hspace{0.1em}]},  Hahn, H.~K. and Mori, K., eds.,
  {\bf 10950},  53, SPIE (mar 2019).

\bibitem{Mabu2017}
Mabu, S., Obayashi, M., Kuremoto, T., Hashimoto, N., Hirano, Y., and Kido, S.,
  ``{Unsupervised class labeling of diffuse lung diseases using frequent
  attribute patterns},'' {\em International Journal of Computer Assisted
  Radiology and Surgery}~{\bf 12},  519--528 (mar 2017).

\bibitem{Dmitriev_2019_CVPR}
Dmitriev, K. and Kaufman, A.~E., ``{Learning Multi-Class Segmentations From
  Single-Class Datasets},'' in [{\em The IEEE Conference on Computer Vision and
  Pattern Recognition (CVPR)}{\nolinebreak\hspace{0.1em}]},  (2019).

\bibitem{Kong2019}
Kong, F., Chen, C., Huang, B., Collins, L.~M., Bradbury, K., and Malof, J.~M.,
  ``{Training a single multi-class convolutional segmentation network using
  multiple datasets with heterogeneous labels: preliminary results},'' in [{\em
  IGARSS 2019 - 2019 IEEE International Geoscience and Remote Sensing
  Symposium}{\nolinebreak\hspace{0.1em}]},   3903--3906, IEEE (jul 2019).

\bibitem{Ronneberger2015}
Ronneberger, O., Fischer, P., and Brox, T., ``{U-Net: Convolutional Networks
  for Biomedical Image Segmentation},'' in [{\em Medical Image Computing and
  Computer-Assisted Intervention -- MICCAI}{\nolinebreak\hspace{0.1em}]},
  234--241, Springer International Publishing (2015).

\end{thebibliography}

\end{document}